\documentclass[11pt,a4paper,reqno]{article}
\usepackage{mysty}
\usepackage{fullpage}

\usepackage[thinlines]{easytable}

\SetLabelAlign{parright}{\parbox[t]{\labelwidth}{\raggedleft{#1}}}
\setlist[description]{style=multiline,topsep=4pt,align=parright}

\makeatletter
\let\reftagform@=\tagform@
\def\tagform@#1{\maketag@@@{(\ignorespaces\textcolor{black}{#1}\unskip\@@italiccorr)}}
\newcommand{\iref}[1]{\textup{\reftagform@{\tcr{\ref{#1}}}}}
\makeatother

\newenvironment{affiliations}{%
    \setcounter{enumi}{1}%
    \setlength{\parindent}{0in}%
    \slshape\sloppy%
    \begin{list}{\upshape$^{\arabic{enumi}}$}{%
        \usecounter{enumi}%
        \setlength{\leftmargin}{0in}%
        \setlength{\topsep}{0in}%
        \setlength{\labelsep}{0in}%
        \setlength{\labelwidth}{0in}%
        \setlength{\listparindent}{0in}%
        \setlength{\itemsep}{0ex}%
        \setlength{\parsep}{0in}%
        }
    }{\end{list}\par\vspace{12pt}}

\begin{document}

\title{A New Framework for Determination of Excitatory and Inhibitory Conductances Using Somatic Clamp}
\author{Songting Li$^{1}$, Xiaohui Zhang$^{2}$, Douglas Zhou$^{3,*}$, \& David Cai$^{1,3,4,*}$}

\date{}
\maketitle
\begin{affiliations}
 \item Courant Institute of Mathematical Sciences and Center for Neural Science, New York University, New York, NY, United States of America
 \item State Key Laboratory of Cognitive Neuroscience and Learning, IDG/McGovern Institute for Brain Research, Beijing Normal University, Beijing, China
 \item School of Mathematical Sciences, MOE-LSC, and Institute of Natural Sciences, Shanghai Jiao Tong University, Shanghai, China
 \item NYUAD Institute, New York University Abu Dhabi, Abu Dhabi, United Arab Emirates\\
\end{affiliations}

\begin{abstract}
The interaction between excitation and inhibition is crucial for brain computation. To understand synaptic mechanisms underlying brain function, it is important to separate excitatory and inhibitory inputs to a target neuron. In the traditional method, after applying somatic current or voltage clamp, the excitatory and inhibitory conductances are determined from the synaptic current-voltage (I-V) relation --- the slope corresponds to the total conductance and the intercept corresponds to the reversal current. Because of the space clamp effect, the measured conductance in general deviates substantially from the local conductance on the dendrite. Therefore, the interpretation of the conductance measured by the traditional method remains to be clarified. In this work, based on the investigation of an idealized ball-and-stick neuron model and a biologically realistic pyramidal neuron model, we first demonstrate both analytically and numerically that the conductance determined by the traditional method has no clear biological interpretation due to the neglect of a nonlinear interaction between the clamp current and the synaptic current across the spatial dendrites. As a consequence, the traditional method can induce an arbitrarily large error of conductance measurement, sometimes even leads to unphysically negative conductance. To circumvent the difficulty of elucidating synaptic impact on neuronal computation using the traditional method, we then propose a framework to determine the effective conductance that reflects directly the functional impact of synaptic inputs on action potential initiation and thereby neuronal information processing. Our framework has been further verified in realistic neuron simulations, thus greatly improves upon the traditional approach by providing a reliable and accurate assessment of the role of synaptic activity in neuronal computation.
\end{abstract}

\section*{Introduction}
The interplay between excitation and inhibition gives rise to rich functions of the brain, for instance, to stabilize and shape neural activities \cite{wehr2003balanced,turrigiano2004homeostatic}, to enhance feature selectivity in sensory neurons \cite{priebe2005direction,xing2011untuned,ye2010synaptic}, and to modulate neural oscillations \cite{buzsaki2012mechanisms,economo2012membrane}. In the meantime, imbalance between excitation and inhibition can induce various neuropsychiatric diseases such as schizophrenia \cite{yizhar2011neocortical,lisman2012excitation}. In order to understand synaptic mechanisms underlying the brain function, a fundamental approach is to quantify pure excitatory and inhibitory components received simultaneously by a target neuron in a neuronal network in the brain. Among electrophysiological recording techniques, somatic current clamp and voltage clamp have become a popular choice to measure excitatory and inhibitory conductances both \emph{in vitro} and \emph{in vivo} studies over the last thirty years \cite{monier2008vitro}. For instance, current clamp has been applied in studies of various brain areas, such as the visual cortex \cite{anderson2000orientation,anderson2001membrane,hirsch1998synaptic,marino2005invariant,priebe2005direction,priebe2006mechanisms}, the barrel cortex \cite{higley2006balanced,wilent2004synaptic,wilent2005dynamics}, and the prefrontal cortex \cite{haider2006neocortical}. Meanwhile, voltage clamp has also been applied in studies of the visual cortex \cite{borg1996voltage,borg1998visual,le2006homeostatic,atallah2012parvalbumin}, the auditory cortex \cite{zhang2003topography,wehr2003balanced,wehr2005synaptic,ye2010synaptic}, the prefrontal cortex \cite{shu2003turning,haider2006neocortical}, and the somatosensory cortex \cite{cruikshank2007synaptic}.

To reveal the quantitative information of excitatory and inhibitory conductances, the traditional method to process data collected under somatic clamp mode is summarized as follows. In this approach, a neuron is viewed as a \emph{point} neuron with its membrane potential dynamics governed by \cite{dayan2001theoretical}
\begin{equation}
\label{eqn:point}
C\frac{dV}{dt}=-G_{L}V-G_{E}(V-\varepsilon_{E})-G_{I}(V-\varepsilon_{I})+I_{\textrm{inj}},
\end{equation}
where $C$ is the membrane capacitance, $V$ is the membrane potential measured at the soma, $G_{L}$, $G_{E}$, and $G_{I}$ are the leak, excitatory, and inhibitory conductances, respectively, $\varepsilon_{E}$ and $\varepsilon_{I}$ are the excitatory and inhibitory reversal potentials, respectively, and $I_{\textrm{inj}}$ is the externally injected current from the somatic clamp. Here all potentials are relative to the resting potential. Based on the point neuron assumption (\ref{eqn:point}), by clamping the somatic current $I_{\textrm{inj}}$ or voltage $V$ at different levels, one can record the corresponding synaptic current
\begin{equation}
I_{\textrm{syn}}=C\frac{dV}{dt}+G_{L}V-I_{\textrm{inj}}
\label{eqn:Isyn}
\end{equation}
obtained by measuring the intracellular trace of somatic membrane potential $V$ (under current clamp mode) or the injected clamp current $I_{\textrm{inj}}$ (under voltage clamp mode) and thereby obtain the linear synaptic current-voltage (I-V) relation. An important assumption in Eq. \ref{eqn:point} is that $G_{E}$ and $G_{I}$ are independent of $I_{\textrm{inj}}$ in this \emph{point} neuron. Under this assumption, the excitatory and inhibitory conductances can then be solved for from the slope and the intercept of the I-V line by casting $-G_{E}(V-\varepsilon_{E})-G_{I}(V-\varepsilon_{I})$ as $I_{\textrm{syn}}$ to obtain $I_{\textrm{syn}}=-kV+b$, where the slope $k$ equals the total conductance defined as the direct sum of the excitatory and inhibitory conductances, and the intercept $b$ equals the reversal current defined as the weighted sum of the excitatory and inhibitory conductances, i.e.,
\begin{equation}
\label{eqn:k1}
k=G_{E}+G_{I},
\end{equation}
\begin{equation}
\label{eqn:b1}
b=G_{E}\varepsilon_{E}+G_{I}\varepsilon_{I}.
\end{equation}

Despite the extensive application of current and voltage clamps to extract excitatory and inhibitory conductances, various important issues related to the validity of the above approach remain to be clarified. As pointed out by theoretical and experimental studies, the voltage distribution across an entire neuron can be highly nonuniform \cite{magee2000dendritic,spruston2008pyramidal}, and a somatic clamp can only exert a limited control of the membrane potential across the dendritic arbor \cite{williams2008direct,poleg2011imperfect}. Therefore, it is important to address the crucial question in which sense a neuron can be viewed as a point as described by Eq. \ref{eqn:point}. In addition, due to the space clamp effect, the conductance measured using a somatic clamp can significantly deviate from the local synaptic conductance on the dendrite \cite{williams2008direct}, sometimes could even yield an unphysically negative value \cite{williams2008direct}. Therefore, it is necessary to carry out a critical assessment of the traditional method of determining conductance.

To address the validity of the point neuron assumption in the traditional method, we begin with the analysis of an idealized passive ball-and-stick model. In the presence of a somatic current injection, say, the clamp current, we demonstrate that the spatial-dependent ball-and-stick model can asymptotically reduce to a point-neuron model to describe the dynamics of the \emph{somatic} membrane potential. In light of this, we consider the soma rather than the entire neuron as a point and introduce the concept of effective conductance, which is defined by Ohm's law as the ratio of the synaptic current arriving at the soma $I^{(0)}_{\textrm{syn}}$ to the driving force (difference between the reversal potential $\varepsilon$ and the somatic membrane potential $V$) in the presence of either excitatory or inhibitory input on the dendrite, i.e., $G^{\textrm{eff}}=\frac{I^{(0)}_{\textrm{syn}}}{\varepsilon-V}$. we emphasize that, in order to distinguish from the synaptic current measured using current or voltage clamp below, here $I^{(0)}_{\textrm{syn}}$ with the superscript ``0'' is the synaptic current in the absence of any injected current $I_{\textrm{inj}}$. As will be demonstrated below, our defined effective conductance is a proportional indicator of the local synaptic conductance and reflects directly the functional impact of synaptic inputs on action potential initiation, hence neuronal information coding.

Our theoretical analysis further demonstrates that it is invalid to simply replace the synaptic conductance in the traditional point neuron model (\ref{eqn:point}) with the effective conductance, and the subthreshold dynamics of the neuron (\ref{eqn:point}) should be corrected by the following dynamics
\begin{equation}
C\frac{dV}{dt}=-G_{L}V-G^{\textrm{inj}}_{E}(V-\varepsilon_{E})-G^{\textrm{inj}}_{I}(V-\varepsilon_{I})+I_{\textrm{inj}}, \label{eqn:point_2}
\end{equation}
where, due to the nonlinear interaction of the injected somatic clamp current with the synaptic current from the dendrites, $G^{\textrm{inj}}_{E}$ and $G^{\textrm{inj}}_{I}$ depend on the injected current $I_{\textrm{inj}}$ to the soma by the clamp. It is only under the condition $I_{\textrm{inj}}=0$ that the effective conductance $G^{\textrm{eff}}_{E}=G^{\textrm{inj}}_{E} (I_{\textrm{inj}}=0)$, $G^{\textrm{eff}}_{I}=G^{\textrm{inj}}_{I} (I_{\textrm{inj}}=0)$, in contrast to the traditional method, in which the conductance is assumed to be independent of the injected current. Therefore, the conductance determined by the traditional method is close to neither the local conductance on the dendrite nor the effective conductance at the soma. Indeed it has no clear biological interpretation.

By applying the somatic clamp, one is purported to measure the effective conductance rather than the local conductance, because the clamp can control the current and voltage sufficiently well only at the soma and its nearby locations. To overcome the inherent difficulty of determining the effective conductance using the traditional method, we present a new method derived from our analysis. Notwithstanding the nonlinear interaction between the synaptic and injected currents, we can obtain the linear I-V relation by changing the injected clamp current at multiple levels of magnitude. Our analysis further shows that the nonlinear current interaction causes the slope of the I-V line to deviate greatly from the sum of two conductances (Eq. \ref{eqn:k1}), however, the intercept of the I-V line remains a good approximation to the reversal current. Therefore, the effective excitatory and inhibitory conductances can be solved from the intercepts (Eq. \ref{eqn:b1}) by varying excitatory or inhibitory reversal potentials at different levels. Finally, we verify our proposed method by employing numerical simulations of the ball-and-stick model and a biologically realistic pyramidal neuron model of complex dendritic morphology and broad ionic channel distribution. In general, our method greatly improves upon the traditional approach in current or voltage clamp by providing a more reliable and accurate assessment of synaptic impact on neuronal computation.

\section*{Methods}
\emph{The ball-and-stick neuron model}\\
We consider an idealized passive ball-and-stick neuron model whose isotropic spherical soma is connected to an unbranched cylindric dendrite with finite length and diameter. The spatiotemporal dynamics of the membrane potential $v(x,t)$ along the dendritic cable is governed by \cite{tuckwell1988introduction,dayan2001theoretical}
\begin{equation}
c\frac{\partial v}{\partial t}=-g_{L}v-{\underset{q=E,I}{\sum }}g_{q}(v-\varepsilon _{q})+I_{\textrm{inj}}+\frac{d}{4r_{a}}\frac{\partial ^{2}v}{\partial x^{2}},
\label{eqn:cable1}
\end{equation}
where $c$ is the membrane capacitance density, $g_{L}$ is the leak conductance density, $g_{E}$ and $g_{I}$ are excitatory and inhibitory conductance densities, respectively, $\varepsilon_{E}$ and $\varepsilon_{I}$ are excitatory and inhibitory reversal potentials, respectively, $I_{\textrm{inj}}$ is externally injected current density, $d$ is the dendritic diameter, and $r_{a}$ is the axial resistivity. Here, all potentials are measured relative to the resting potential.

When excitatory and inhibitory inputs are elicited at dendritic sites, we have the synaptic input
\begin{equation}
g_{q}=\overset{M_{q}}{\underset{i=1}{\sum }}\overset{\infty}{\underset{j=1}{\sum }}f^{ij}_{q}u_{q}(t-t^{ij}_{q})\delta (x-x_{q}^{i}), \nonumber
\end{equation}
where $q=E,I$. $M_{E}$ ($M_{I}$) is the number of dendritic sites for excitatory (inhibitory) inputs. For an individual synaptic input of type $q$, $f^{ij}_{q}$ is the input strength of the $j$-th input at the $i$-th location $x^{i}_{q}$ with its arrival time $t^{ij}_{q}$. Here $u_{q}(t)$ is the unitary conductance density pulse modeled as $u_{q}(t)=N_{q}( e^{-\frac{t}{\sigma _{qd}}}-e^{-\frac{t}{\sigma _{qr}}})\Theta(t)$ with the peak value normalized to unity by the normalization factor $N_{q}=\Big[\Big(\frac{\sigma_{qr}}{\sigma_{qd}}\Big)^\frac{\sigma_{qr}}{\sigma_{qd}-\sigma_{qr}}-
\Big(\frac{\sigma_{qr}}{\sigma_{qd}}\Big)^\frac{\sigma_{qd}}{\sigma_{qd}-\sigma_{qr}}\Big]^{-1}$, where $\Theta(t)$ is the Heaviside function, $\sigma _{qr}$ and $\sigma _{qd}$ are the rise and decay time constants of individual synaptic conductance, respectively \cite{dayan2001theoretical}.

The assumption that one end of the dendrite is sealed yields
\begin{equation}
\left. \frac{\partial v}{\partial x}\right\vert _{x=l}=0,
\label{eqn:bc1}
\end{equation}
where $l$ is the dendritic length. At the other end that connects to the soma, the law of current conservation gives rise to
\begin{equation}
c\frac{\partial v(0,t)}{\partial t}=-g_{L}v(0,t)+\frac{\pi d^{2}}{4Sr_{a}}%
\left. \frac{\partial v}{\partial x}\right\vert _{x=0},
\label{eqn:bc2}
\end{equation}
where $S$ is the somatic surface area. Eqs. \ref{eqn:bc1} and \ref{eqn:bc2} constitute the boundary conditions of the cable model (\ref{eqn:cable1}). Before the arrival of synaptic inputs, the neuron stays at the resting state with the initial condition set as $v(x,0)=0$.

For the numerical simulation of the ball-and-stick neuron (\ref{eqn:cable1}), the Crank-Nicolson method \cite{thomas2013numerical} is used with time step 0.1 ms and space grid size of 1 $\mu$m. Parameters in our simulation are within the physiological regime \cite{koch2004biophysics,hao2009arithmetic} with $c=1$ $\mu$F$\cdot$cm$^{-2}$, $g_{L}=0.05$ mS$\cdot$cm$^{-2}$, $\protect\varepsilon_{E}=70$ mV, $\protect\varepsilon_{I}=-10$ mV, $S=2.83\times10^3$ $\mu$m$^{2}$, $r_{a}=100$ $\Omega$$\cdot$cm, $l=600$ $\mu$m, $d=1$ $\mu$m, $\sigma_{Er}=5$ ms, $\protect\sigma_{Ed}=7.8$ ms, $\protect\sigma_{Ir}=6$ ms, and $\protect\sigma_{Id}=18$ ms. The time constants here are chosen to be consistent with the conductance inputs in the experiment \cite{hao2009arithmetic,li2014bilinearity}.\\

\noindent \emph{The realistic neuron model}\\
The realistic pyramidal neuron model is adapted from our previous studies of dendritic integration \cite{hao2009arithmetic,li2014bilinearity,li2015analysis} (see Ref. \cite{hao2009arithmetic} for details). The morphology of the reconstructed pyramidal neuron, which contains 200 compartments, is obtained from the Duke-Southampton Archive of neuronal morphology \cite{cannon1998line}. The passive cable properties and density distribution of active conductances in the model neuron are based on published experimental data for hippocampal and cortical pyramidal neurons \cite{stuart1998determinants,magee1995characterization,hoffman1997k,migliore1999role,magee1998dendritic,magee2000somatic,andrasfalvy2001distance,smith2003mechanism,nicholson2006distance}. The model also contains AMPA, NMDA, GABA$_{A}$ and GABA$_{B}$ receptors with kinetic properties derived from Refs. \cite{destexhe1994efficient,destexhe1994synthesis,destexhe1998kinetic}. We use the NEURON software Version 7.3 \cite{carnevale2006neuron} to simulate the model with time step 0.1 ms.

\section*{Results}
\textbf{Geometrical reduction}\\
The traditional method for extracting excitatory and inhibitory conductances is based on a crucial assumption that a neuron can be considered as a point with its membrane potential dynamics described by Eq. \ref{eqn:point}. However, because of the highly nonuniform distribution of membrane potential across a neuron \cite{magee2000dendritic,spruston2008pyramidal}, the entire neuron is not electrically compact, thus cannot be modeled as a point. On the other hand, in experiment, it has been shown that the membrane potential dynamics of the soma of a neuron (relative to the resting potential) can be well captured by a point leaky integrator \cite{carandini1996spike,badel2008dynamic}
\begin{equation}
\label{eqn:leaky}
C\frac{dV}{dt}=-G_{L}V+I_{\textrm{inj}}
\end{equation}
in response to the externally injected current $I_{\textrm{inj}}$ into the soma. Therefore, one should model the soma rather than the entire neuron as a point. However, there is a lack of theoretical demonstration of how to obtain a point characterization of the soma from a spatial-dependent neuron model in general.

For a large class of neurons, the tree-like passive dendrites can be shown to be mathematically equivalent to a single cylindric cable \cite{rall1962theory}. To demonstrate the validity of the point characterization (Eq. \ref{eqn:leaky}), without loss of generality, we therefore start with the ball-and-stick neuron introduced in the \emph{Materials and Methods} section. Given a current pulse input $I_{\delta}$ at the \emph{soma}, the ball-and-stick model
\begin{equation}
c\frac{\partial v}{\partial t}=-g_{L}v+I_{\delta}+\frac{d}{4r_{a}}\frac{\partial ^{2}v}{\partial x^{2}} \nonumber
\end{equation}
possesses the following response kernel (Green's function) that captures the \emph{somatic} membrane potential response $V(t)\equiv v(0,t)$ \cite{tuckwell1988introduction,li2014bilinearity,li2015analysis},
\begin{equation}
\Gamma(t)=\sum\limits_{n}H_{n}e^{-k_{n}t},
\label{eqn:green}
\end{equation}
where constant coefficients $H_{n}$ and time constants $k_{n}$ are determined by the geometry and biophysics of the passive neuron. Asymptotically, the response kernel can be well approximated by its leading order with a single time constant, i.e.,
\begin{equation}
\label{eqn:asy1}
\Gamma(t)\sim H_{0}e^{-k_{0}t},
\end{equation}
where $k_{0}=g_{L}/c$, $H_{0}=[\gamma/(\gamma\lambda+1)\pi d](4r_{a}/c^{2}d)^{-1/2}$, with $\gamma =\left( \pi d^{2}/2S\right) \left( r_{a}d\right)^{-1/2}$ and $\lambda =l\sqrt{4r_{a}/d}$. Note that Eq. \ref{eqn:asy1} is precisely the response kernel for the point-neuron model (Eq. \ref{eqn:leaky}) with the following relation linking the parameters in the point-neuron model and those in the ball-and-stick neuron, $C=1/H_{0}$, and $G_{L}=k_{0}/H_{0}$. For any time-dependent somatic current input, the somatic response of the ball-and-stick neuron can then be described by the convolution of the response kernel (\ref{eqn:asy1}) of the point-neuron model with the input, thus reducing the somatic membrane potential dynamics of the ball-and-stick neuron with somatic input to its equivalent dynamics of the point-neuron (\ref{eqn:leaky}). Our asymptotic analysis has been further verified numerically to demonstrate that the point-neuron characterization is sufficiently accurate to represent the somatic membrane potential dynamics of the ball-and-stick neuron.\\

\noindent\textbf{Effective conductance}\\
By considering the soma rather than the entire neuron as an electrically compact point, the concept of effective conductance naturally arises following the Ohm's law, which casts, say, the effective excitatory conductance, in the form
\begin{equation}
\label{eqn:synaptic_G}
G^{\textrm{eff}}_{E}=\frac{I^{(0)}_{\textrm{syn}}}{\varepsilon_{E}-V},
\end{equation}
where $I^{(0)}_{\textrm{syn}}$ is the synaptic current arriving at the soma in the absence of any injected current to the soma, $\varepsilon_{E}$ is the excitatory reversal potential relative to the resting potential, and $V(t)$ is the somatic membrane potential change in response to an excitatory synaptic input from its dendrite. The synaptic current arriving at the soma is determined from the point-neuron model by
\begin{equation}
\label{eqn:synaptic_I}
I^{(0)}_{\textrm{syn}}=C\frac{dV}{dt}+G_{L}V,
\end{equation}
which will be referred to as the effective synaptic current below (note that $I_{\textrm{inj}}=0$ in Eq. \ref{eqn:synaptic_I}). A similar definition holds for the effective inhibitory conductance. The effective synaptic current measured at the soma can be significantly different from the synaptic current measured at the synapse localized on the dendrite. This arises because the local synaptic current induced at the synapse will be filtered by the dendritic cable property and further modified by the interaction with active ion channels along dendrites before reaching the soma \cite{stuart1998determinants,williams2002dependence,magee2000dendritic,hille2001ion,london2005dendritic}.

Conceptually, it is evident that the effective conductance at the soma and local conductance on the dendrite are rather different. As demonstrated by our numerical simulation of the ball-and-stick neuron, there is indeed a significant quantitative difference between them. Shown in Figures 1A and 1D are the numerically measured effective conductances, which are significantly smaller than the local conductances upon an excitatory or inhibitory Poisson input of rate 150 Hz on the dendrite at the location 420 $\mu$m away from the soma. The effective excitatory and inhibitory conductances are found to be strongly correlated with the corresponding local conductances with correlation coefficient $\rho=0.95$ and $\rho=0.99$ respectively, as shown in Figures 1B and 1E. This correlation suggests that the effective conductance is indeed a good indicator to reflect the synaptic activity on the dendrites. It is reasonable to expect that the effective conductance decreases gradually with the increase of distance between the input location and the soma. This is confirmed in Figures 1C and 1F. Therefore, a strong input at a site on a distal dendrite and a weak input at a site on the proximal dendrite may induce a membrane potential change of similar magnitude at the soma. That is, local conductances on the dendrite can differ greatly for different inputs while the corresponding effective conductances could exert a similar impact on the somatic membrane potential dynamics. In this sense, the effective conductance reflects directly the functional impact of synaptic inputs originated from the dendrite on somatic membrane potential, action potential generation, hence neuronal information coding. In short, the effective conductance plays a central role in quantifying the synaptic influence on neuronal computation.\\

\begin{figure}
  \begin{center}
    \includegraphics[width=1\textwidth]{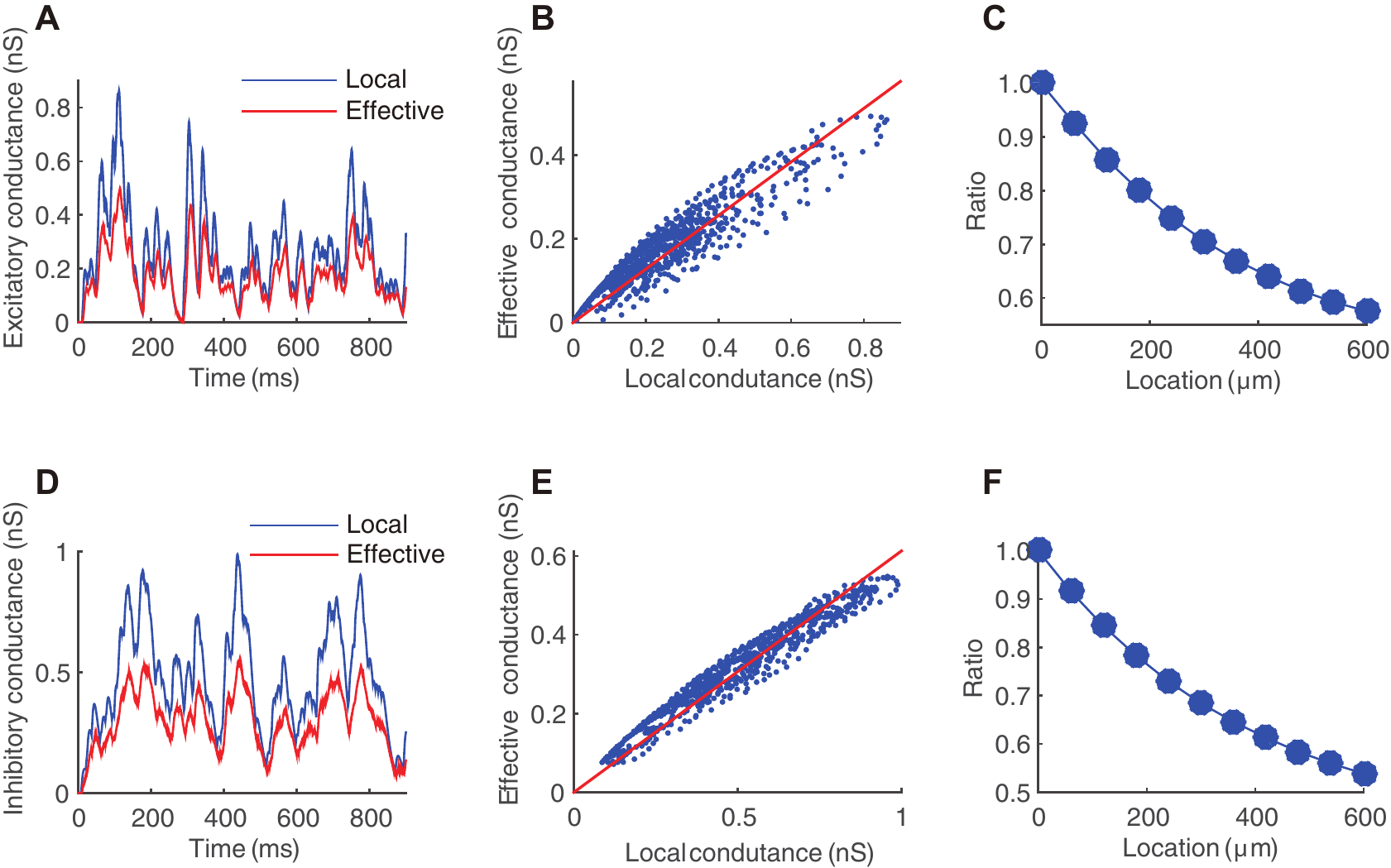}
  \end{center}
  \caption{Properties of effective synaptic conductances. (A), trace of the effective and local excitatory synaptic conductances measured respectively at the soma and at a synapse on the dendrite of the ball-and-stick neuron. An excitatory Poisson input with rate 150 Hz is applied on the dendritic site that is 420 $\mu$m away from the soma. (B), strong correlation between the effective and local synaptic conductance. Each blue dot is sampled at one time point from the corresponding time series in \emph{A}. The red straight line is the linear fitting with correlation coefficient $\rho=0.95$. (C), The dependence of the ratio of the effective to the local excitatory conductance on the distance between the input location and the soma. The ratio of the effective to the local conductance is defined as the slope of the linear fitting in \emph{B}. (D-F) are for the inhibitory case. An inhibitory Poisson input with rate 150 Hz is applied on the dendrite at the location 420 $\mu$m away from the soma.}
\end{figure}

\noindent \textbf{Interaction between clamp current and synaptic current}\\
As is shown in the \emph{Geometrical reduction} subsection, for an injected current to the \emph{soma} of the ball-and-stick neuron, the point-neuron model (Eq. \ref{eqn:leaky}) is quantitatively accurate for describing its somatic membrane potential change. However, contrary to the conventional belief, the point-neuron model (Eq. \ref{eqn:point}) becomes invalid in the presence of both clamp current at the \emph{soma} and synaptic current from the \emph{dendrite}. As revealed in our analysis below, because of nonlinearity in the interaction between the somatic clamp current and the synaptic current, no longer can the traditional point-neuron model (Eq. \ref{eqn:point}) provide a conceptually correct and quantitatively accurate description of the true somatic voltage dynamics.

In the point-neuron model (Eq. \ref{eqn:point}), all the synaptic currents as well as the injected current are assumed to be summed linearly at the soma. However, the synaptic current is voltage-dependent and the injected current at the soma can change the membrane potential on the dendrite, thus resulting in nonlinear interactions between the injected current and the synaptic current. Therefore, these two currents can no longer be summed directly at the soma. We now present the detailed analysis of the origin of the nonlinear interaction using the ball-and-stick neuron model. For the sake of illustration, we discuss the case of the excitatory input. At the time $t=0$, given an excitatory input at the dendrite $x=x_{E}$ and a constant injected current $I_{\textrm{inj}}$ at the soma $x=0$, we have
\begin{equation}
c\frac{\partial v}{\partial t}=-g_{L}v-
f_{E}u_{E}(t)\delta (x-x_{E})(v-\varepsilon _{E})+\frac{I_{\textrm{inj}}}{\pi d}\Theta(t)\delta (x-0)+\frac{d}{4r_{a}}\frac{%
\partial ^{2}v}{\partial x^{2}}. \nonumber
\end{equation}
In the physiological regime, in general, an individual synaptic input can only effect a small change of somatic membrane potential. This gives rise to an asymptotic expansion of the somatic response $V(t)\equiv v(0,t)$ with respect to the input strength $f_{E}$ \cite{li2014bilinearity,li2015analysis},
\begin{equation}
V(t)=\overset{\infty}{\underset{k=0}{\sum }}f_{E}^{k}V_{k}(t).
\label{eqn:asyV}
\end{equation}
Using the Green's function method, the zeroth and first order solution at the soma can be cast into the following form
\begin{equation}
V_{0}=a_{0}I_{\textrm{inj}},\ \ \ \ \ V_{1}=a_{1}I_{\textrm{inj}}+b_{1}, \nonumber
\end{equation}
where the coefficients are $a_{0}=\Gamma(0,0,t)*\Theta(t)/ \pi d$, $a_{1}=-\Gamma(0,x_{E},t)*[u_{E}(t)\cdot \Gamma(x_{E},0,t)*\Theta(t)]/\pi d$, and $ b_{1}=\Gamma(0,x_{E},t)*[\varepsilon_{E} \cdot u_{E}(t)]$. Here $\Gamma(x,y,t)$ is the response kernel (Green's function) of the cable equation whose explicit expression has been derived in Refs. \cite{li2014bilinearity,li2015analysis}, $\Theta(t)$ is the heaviside function, ``$*$'' is the temporal convolution, and ``$\cdot$'' is the multiplication. Using the synaptic current as measured by $I_{\textrm{syn}}=C\frac{dV}{dt}+G_{L}V-I_{\textrm{inj}}$, together with Eq. \ref{eqn:asyV}, we can readily derive an expression for the synaptic current at the soma
\begin{eqnarray}
\label{interaction}
I_{\textrm{syn}}=f_{E}(Ca^{'}_{1}+G_{L}a_{1})I_{\textrm{inj}}+f_{E}(Cb^{'}_{1}+G_{L}b_{1})
\end{eqnarray}
to the first order of $f_E$, where the prime stands for the derivative with respect to time. In the derivation the equality $CV^{'}_{0}+G_{L}V_{0}-I_{\textrm{inj}}=0$ is used because $V_{0}$ is the zeroth order membrane potential change, which responds to the injected current $I_{\textrm{inj}}$ only (see Eq. \ref{eqn:leaky}). The first term in Eq. \ref{interaction} of the synaptic current describes the interaction between the injected current and the current from the dendrite, and the second term describes the effective synaptic current $I^{(0)}_{\textrm{syn}}$ when measured in the absence of the injected current. Using Eq. \ref{interaction}, an expression for the excitatory conductance $G^{\textrm{inj}}_{E}=\frac{I_{\textrm{syn}}}{\varepsilon_{E}-V}$ under current injection can be obtained to the first order of $f_{E}$ as
\begin{eqnarray}
\label{eff}
G^{\textrm{inj}}_{E}=\frac{f_{E}(Ca^{'}_{1}+G_{L}a_{1})I_{\textrm{inj}}+I^{(0)}_{\textrm{syn}}}{\varepsilon_{E}-a_{0}I_{\textrm{inj}}},
\end{eqnarray}
from which it is evident that $G^{\textrm{inj}}_{E}$ depends on the injected current $I_{\textrm{inj}}$. The superscript of $G^{\textrm{inj}}_{E}$ emphasizes the fact that this excitatory conductance is measured in the presence of an injected current. In the conventional approach, one assumes that $G_{E}$ in Eq. \ref{eqn:point} is not affected by the injected current and asserts that the injected current merely induces a somatic membrane potential change to yield the corresponding synaptic current $\Delta I_{\textrm{syn}}=G_{E}(\varepsilon_{E}-\Delta V)$ with an $I_{\textrm{inj}}$-independent $G_{E}$. Therefore, the effect of the interaction between the two currents is not accounted for. It can be clearly seen from Eq. \ref{eff} that the value of excitatory conductance $G^{\textrm{inj}}_{E}$ is modified by the injected current in contrast to the case of the presence of a purely excitatory synaptic current. Similarly, the injected current will interact with an inhibitory synaptic input, resulting in a modified value of the inhibitory conductance $G^{\textrm{inj}}_{I}$ in comparison with the value for a purely inhibitory synaptic current case. It is only under the condition $I_{\textrm{inj}}=0$ that the effective conductances $G^{\textrm{eff}}_{E}=G^{\textrm{inj}}_{E} (I_{\textrm{inj}}=0)$, and $G^{\textrm{eff}}_{I}=G^{\textrm{inj}}_{I} (I_{\textrm{inj}}=0)$.

We next solve the ball-and-stick model numerically to further confirm the validity of the asymptotic results above. Our numerical results again affirm the important role of the interaction between injected current and synaptic current in the determination of conductance value. Given an individual excitatory synaptic input at a dendritic site away from the soma together with an injected constant current at the soma of the ball-and-stick neuron, we can numerically obtain the somatic membrane potential of the ball-and-stick neuron $V$ and measure its synaptic current $I_{\textrm{syn}}$ based on the point neuron model, $I_{\textrm{syn}}=C\frac{dV}{dt}+G_{L}V-I_{\textrm{inj}}.$ By varying the strength of the injected constant current while maintaining the strength of the local synaptic input on the dendritic site, we can obtain synaptic current $I_{\textrm{syn}}$ of various amplitudes. An example is displayed in Figure 2A for particular realizations of this procedure. As has been predicted by our theoretical analysis (Eq. \ref{interaction}), the linear dependence of the measured peak amplitude of the synaptic current $I_{\textrm{syn}}$ on the injected constant current is confirmed in Figure 2B. We are now ready to determine the excitatory conductance $G^{\textrm{inj}}_{E}$ under current injection using $G^{\textrm{inj}}_{E}=\frac{I_{\textrm{syn}}}{\varepsilon_{E}-V}$. Contrary to the common belief that the synaptic conductance measured at the soma is independent of the injected current, here the excitatory conductance strongly depends on the injected current. As shown in Figures 2C-2D, the difference can range from 7\% to a substantial 36\% between the peak excitatory conductance $G^{\textrm{inj}}_{E}$ in the presence of the injected constant current and $G^{\textrm{eff}}_{E}$ in the absence of the injected current. In fact, the difference can be arbitrarily large as the magnitude of the injected current further increases. Similarly, as shown in Figures 2E-2H, given an individual inhibitory synaptic input at a dendritic location away from the soma together with an injected constant current at the soma of the ball-and-stick neuron, the inhibitory conductance $G^{\textrm{inj}}_{I}$ also strongly depends on the injected currents. The difference can be rather significant, ranging from 17\% to 153\% between the peak inhibitory conductance $G^{\textrm{inj}}_{I}$ in the presence of the injected constant current and $G^{\textrm{eff}}_{I}$ in the absence of the injected current. As shown in Figure 2H, a negative value of the inhibitory conductance can even be observed under certain magnitude of the injected current. It is demonstrated in Figures 2D and 2H that the nonlinear dependence of the conductance $G^{\textrm{inj}}$ on the injected constant current can be accurately predicted by our analysis (Eq. \ref{eff}).

\begin{figure}
  \begin{center}
    \includegraphics[width=1\textwidth]{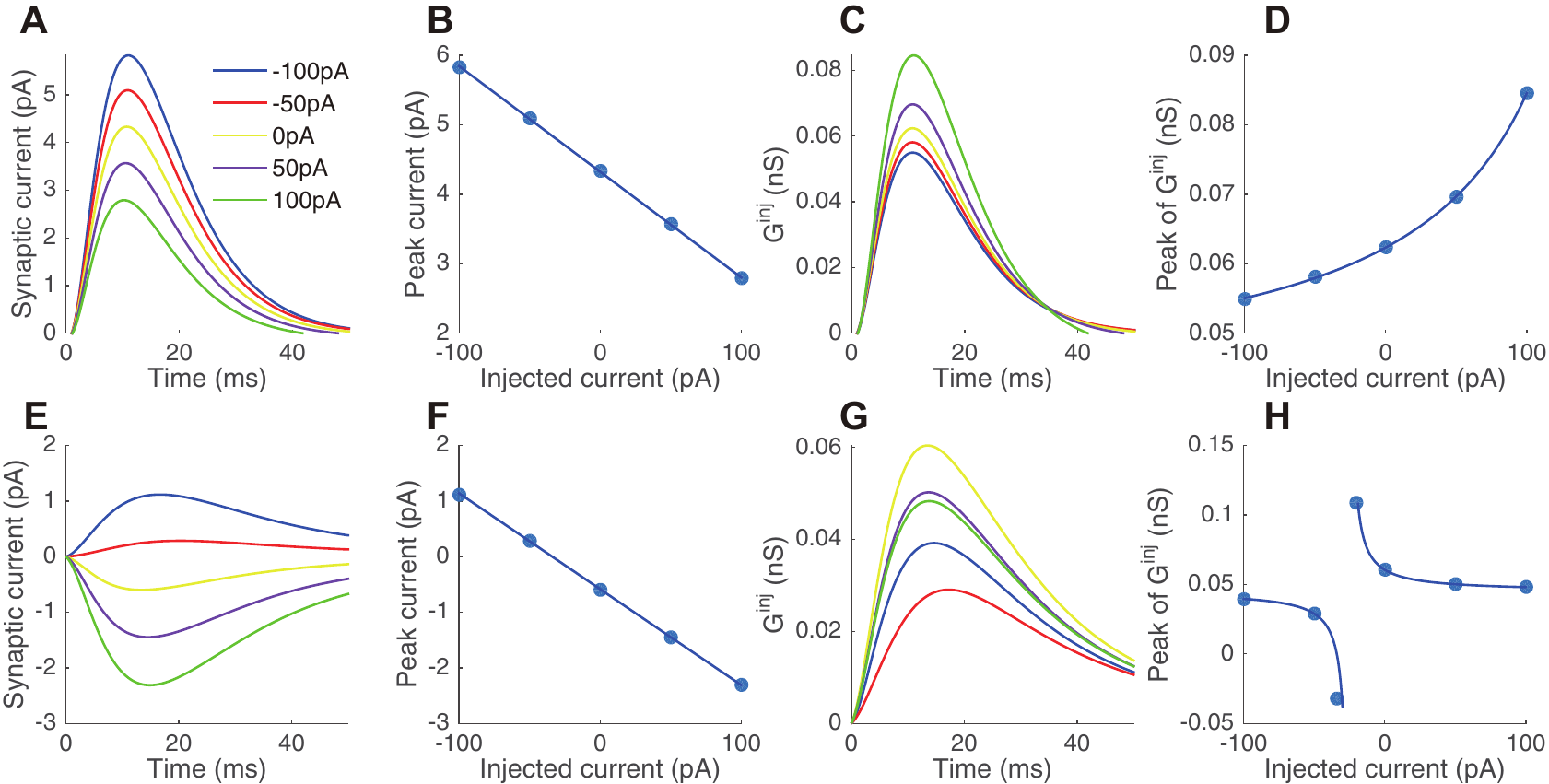}
  \end{center}
  \caption{Determination of the conductance taking into account of the interaction between the injected current and the synaptic current. (A), numerically measured excitatory synaptic current temporal profiles of the point-neuron model in the presence of an injected constant current at the soma of the ball-and-stick neuron. Color indicates the strength of the injected current as labeled on the legend. (B), linear dependence of the peak amplitude of the excitatory synaptic current on the injected current as predicted by Eq. \ref{interaction}. (C), the temporal profile of $G^{\textrm{inj}}_{E}$ numerically measured using the excitatory synaptic currents in \emph{A}. The yellow curve corresponds to $G^{\textrm{eff}}_{E}$. (D), nonlinear dependence of the peak amplitude of $G^{\textrm{inj}}_{E}$ on the injected constant current. (E-H) are the corresponding case for the inhibitory synaptic input. An individual excitatory pulse input is given at the location on the dendrite 420 $\mu$m away from the soma. An individual inhibitory pulse input is given at the location on the dendrite 300 $\mu$m away from the soma. \emph{A}, \emph{C}, \emph{E} and \emph{G} share the same color-coded legend. In \emph{B}, \emph{D}, \emph{F} and \emph{H}, dots are values obtained from numerical simulation, and solid curves are the prediction of the first order asymptotic analysis of the ball-and-stick neuron (see text).}
\end{figure}

Because of the nonlinear dependence of $G^{\textrm{inj}}$ on the injected current, no longer can the synaptic current arriving at the soma and the injected current be simply summed linearly in the point neuron model. As a consequence, the point neuron model should be reformulated by
\begin{equation}
\label{eqn:newpoint}
C\frac{dV}{dt}=-G_{L}V-G^{\textrm{inj}}_{E}(V-\varepsilon_{E})-G^{\textrm{inj}}_{I}(V-\varepsilon_{I})+I_{\textrm{inj}},
\end{equation}
 in which the conductances $G^{\textrm{inj}}_{E}$ and $G^{\textrm{inj}}_{I}$ are function of $I_{\textrm{inj}}$. It is worthwhile to point out that in the presence of even only one type of the synaptic input, the conductance determined based on the traditional point neuron model already requires a significant correction due to the interaction between the clamp current and the synaptic current.\\

\noindent \textbf{Theoretical analysis of conductance measurement}\\
In general, a neuron receives a mixture of excitatory and inhibitory inputs from neighbouring neurons. As mentioned in the \emph{Introduction} section, the traditional way of extracting the excitatory and inhibitory conductance is to apply the current or voltage clamp technique. After injecting current at the soma with different magnitudes, one can measure the corresponding synaptic current and somatic voltage to obtain thereby a linear I-V relation. By assuming the neuron as an electrically compact point (Eq. \ref{eqn:point}), and the excitatory and inhibitory conductances $G_{E}$ and $G_{I}$ are independent of the injected current, $G_{E}$ and $G_{I}$ can then be determined by solving two equations (Eqs. \ref{eqn:k1} and \ref{eqn:b1}) involving the slope and the intercept of the I-V line. However, as shown in our analysis above, the interaction between the somatic clamp current and the synaptic current arriving at the soma renders the point neuron assumption (\ref{eqn:point}) invalid for measuring the conductance even in the presence of a single type of synaptic input, hence the failure of the traditional method to measure the effective conductance. In the following, we demonstrate that, even if the magnitude of the injected clamp current is sufficiently small, the traditional method can still induce an arbitrarily large error of conductance measurement, sometimes even leads to an unphysically negative conductance.

For the ball-and-stick neuron, in the presence of a pair of excitatory and inhibitory synaptic inputs on the dendritic site $x=x_{E}$ and $x=x_{I}$, respectively, the dynamics of its membrane potential is governed by
\begin{equation}
c\frac{\partial v}{\partial t}=-g_{L}v-\underset{q =E ,I}{\sum }
f_{q}u_{q}(t)\delta (x-x_{q})(v-\varepsilon _{q})+\frac{I_{\textrm{inj}}}{\pi d}\Theta(t)\delta (x-0)+\frac{d}{4r_{a}}\frac{%
\partial ^{2}v}{\partial x^{2}} \nonumber
\end{equation}
when the clamp current $I_{\textrm{inj}}$ is applied at the soma. As mentioned above, within the physiological range, the strength of each individual input $f_{E}$ and $f_{I}$ is usually small, we can expand the somatic membrane potential $V(t)\equiv v(0,t)$ as an asymptotic series with respect to the input strengths $f_{E}$ and $f_{I}$,
\begin{equation}
V(t)=\overset{\infty}{\underset{k=0}{\sum }}\underset{\ \ m+n=k}{\sum }f_{E}^{m}f_{I}^{n}V_{mn}(t). \nonumber
\end{equation}
The Green's function method yields the zeroth and first order solutions
\begin{equation}
V_{00}=a_{00}I_{\textrm{inj}},\ \ \ \ \ V_{10}=a_{10}I_{\textrm{inj}}+b_{10}, \ \ \ \ \ V_{01}=a_{01}I_{\textrm{inj}}+b_{01},\nonumber
\end{equation}
where the coefficients are $a_{00}=\Gamma(0,0,t)*\Theta(t)/ \pi d$, $\ a_{10}=-\Gamma(0,x_{E},t)*[u_{E}(t)\cdot \Gamma(x_{E},0,t)*\Theta(t)]/\pi d$, $\ b_{10}=\Gamma(0,x_{E},t)*[\varepsilon_{E} \cdot u_{E}(t)]$, $\ a_{01}=-\Gamma(0,x_{I},t)*[u_{I}(t)\cdot \Gamma(x_{I},0,t)*\Theta(t)]/\pi d$, and $ b_{01}=\Gamma(0,x_{I},t)*[\varepsilon_{I} \cdot u_{I}(t)]$.

Then, to the first order of input strengths $f_q$ ($q=E, I$), the synaptic current in the point neuron model becomes
\begin{equation}
I_{\textrm{syn}}=[C(f_{E}a^{'}_{10}+f_{I}a^{'}_{01})+G_{L}(f_{E}a_{10}+f_{I}a_{01})]I_{\textrm{inj}}+[C(f_{E}b^{'}_{10}+f_{I}b^{'}_{01})+G_{L}(f_{E}b_{10}+f_{I}b_{01})] \nonumber
\end{equation}
in which the first term describes the interaction between the synaptic current and the injected current and the second term describes the effective synaptic current in the absence of the injected current. The corresponding somatic voltage to the first order of $f_q$ ($q=E, I$) is
\begin{equation}
\label{eqn:sv}
V=(a_{00}+f_{E}a_{10}+f_{I}a_{01})I_{\textrm{inj}}+(f_{E}b_{10}+f_{I}b_{01}). \\
\end{equation}
Similarly, the first term on the right hand side of Eq. \ref{eqn:sv} arises from the interaction between the synaptic current and the injected current, and the second term comes from the effective synaptic current in the absence of the injected current. Because both the membrane potential $V$ and the synaptic current $I_{\textrm{syn}}$ are linearly related to the injected current $I_{\textrm{inj}}$, simple linear algebra yields the linear dependence of the synaptic current on the membrane potential, i.e., $I_{\textrm{syn}}= -kV+b$, which holds for arbitrary $I_{\textrm{inj}}$. The corresponding slope and intercept to the first order of $f_q$ ($q=E, I$) become
\begin{eqnarray}
\label{eqn:slope}
k=-\frac{(Ca^{'}_{10}+G_{L}a_{10})\varepsilon_{E}}{(Cb^{'}_{10}+G_{L}b_{10})a_{00}}G^{\textrm{eff}}_{E}-\frac{(Ca^{'}_{01}+G_{L}a_{01})\varepsilon_{I}}{(Cb^{'}_{01}+G_{L}b_{01})a_{00}}G^{\textrm{eff}}_{I}
\end{eqnarray}
and
\begin{eqnarray}
\label{eqn:intercept}
b=G^{\textrm{eff}}_{E}\varepsilon_{E}+G^{\textrm{eff}}_{I}\varepsilon_{I},
\end{eqnarray}
respectively. Note that the slope here is not equal to the total conductance $G^{\textrm{eff}}_{E}+G^{\textrm{eff}}_{I}$ as in the traditional method. The prefactors of $G^{\textrm{eff}}_{E}$ and $G^{\textrm{eff}}_{I}$ in the slope expression (Eq. \ref{eqn:slope}) can be much smaller than unity \cite{Li2017Gdetermining}, thus leading to the failure of the traditional method (using Eqs. \ref{eqn:k1} and \ref{eqn:b1}) in determining the effective conductances. Apparently, the conductance determined by the traditional method has no clear biological interpretation. It is worthwhile to stress that these prefactors are independent of the magnitude of the injected clamp current. As a result, the measurement error of the effective conductance by the traditional method cannot be eliminated even if the magnitude of the clamp current is sufficiently small. This will be further confirmed by numerical results below. In general, the prefactors are dependent of time and input locations, thus rarely can they be determined in advance without the knowledge of the synaptic location and the corresponding response kernel to the synaptic location --- the situation could become even more complicated when a neuron receives numerous inputs across its dendrites. Importantly, the relation of the intercept and the reversal current, however, can now provide a basis for determining the effective excitatory and inhibitory conductances. Because there are two unknowns $G^{\textrm{eff}}_{E}$ and $G^{\textrm{eff}}_{I}$, we need obtain at least one more intercept value. This can be achieved by varying one of the reversal potentials. For example, we can change the inhibitory reversal potential from $\varepsilon_{I}$ to $\hat{\varepsilon_{I}}$ to obtain a second intercept equation
\begin{equation}
\label{eqn:intercept2}
\hat{b}=G^{\textrm{eff}}_{E}\varepsilon_{E}+G^{\textrm{eff}}_{I}\hat{\varepsilon_{I}},
\end{equation}
thereby, the effective excitatory and inhibitory conductances can be obtained from Eqs. \ref{eqn:intercept} and \ref{eqn:intercept2}. In physiological experiment, to change reversal potential, one needs to effect a change of the extracellular or intracellular fluid environment. From now on, we refer to this new method as the intercept method, and the traditional method as the slope-and-intercept method.\\

\noindent \textbf{Numerical verification of the intercept method}\\
We next perform numerical simulations of the ball-and-stick neuron to demonstrate the validity of the intercept method by contrasting its error with that of the traditional slope-and-intercept method.

Given an individual excitatory pulse input at a dendritic location away from the soma but without the injected current at the soma, we can numerically record the corresponding EPSP at the soma, and using Eq. \ref{eqn:synaptic_G}, determine the value of the effective excitatory conductance pulse from the point-neuron model. A similar procedure can be carried out for the effective inhibitory conductance pulse at the soma in response to an inhibitory pulse input at a location on the dendrite away from the soma. A pair of numerically measured effective excitatory and inhibitory conductance pulses determined this way as displayed in Figures 3A and 4A will be used below as the reference values, against which we evaluate the performance of the intercept method and the slope-and-intercept method. Note that these reference conductances are true effective conductances without the distortion induced by the clamp current.

Because we need to tune the reversal potential to a different value at least once in the intercept method, it is important to verify that the effective excitatory and inhibitory conductances are independent of the change of synaptic reversal potential values. This is indeed the case for the ball-and-stick neuron: The independence of the effective conductance on the synaptic reversal potential can be analytically shown from our asymptotic analysis by setting $I_\textrm{inj}=0$ in Eq. \ref{eff} to the first order of $f_q$ ($q=E, I$). In our simulation, the change of $G^{\textrm{eff}}_{E}$ is less than 5\% in value when the excitatory reversal potential varies from 20 mV to 120 mV, while the change of $G^{\textrm{eff}}_{I}$ is less than 0.7\% in value when the inhibitory reversal potential varies from 0 mV to $-20$ mV. We will change only the inhibitory reversal potential to determine the effective conductance using the intercept method below.

To evaluate the performance of the intercept method in comparison with the slope-and-intercept method, we now apply an injected constant clamp current at the soma and simultaneously elicit the same excitatory and inhibitory pulse inputs as the reference ones, i.e. inputting at the same dendritic location with the same strength. By measuring the membrane potential at the soma, we can use Eq. \ref{eqn:Isyn} to obtain a set of the corresponding synaptic current traces in response to the injected current of different amplitudes. At each moment of time, we observe that the linear relation between the synaptic current and the membrane potential persists as usual. We then determine the value of the excitatory and inhibitory conductance pulses from the linear I-V relation moment by moment. At each time point, by using the slope-and-intercept method, we can obtain one pair of the values of $G_{E}$ and $G_{I}$. Meanwhile, by tuning the inhibitory reversal potential from $-10$ mV to $-20$ mV, and repeating the above procedure, we can obtain a second linear I-V relation. Then using the intercept method, we can determine the pair of the values of $G_{E}$ and $G_{I}$. By repeating the same procedure at different time points, we can determine the temporal profiles of the conductance pulses by both methods for the comparison of conductances in the presence of the current clamp with those of the reference conductance pulses measured in the absence of the current clamp. As demonstrated in Figure 3A, it is evident that the value of the conductance pulse determined from the intercept method is much more accurate than those reconstructed from the slope-and-intercept method, in particular, for inhibitory case. The effective conductance pulses measured using the intercept method have a relatively small error, with the maximum error value of 2\% for excitatory and for inhibitory conductance, in contrast with those determined using the slope-and-intercept method, which yields an error as large as 6\% for excitatory conductance and 35\% for inhibitory conductance. We have also numerically confirmed that the large error of the slope-and-intercept method cannot be significantly reduced even when the magnitude of the injected clamp current decreases to 1\% of the original one.

In the brain, neurons dynamically receive synaptic inputs all the time. This requires us to address the issue of how to determine conductance under this condition. We consider the case of two excitatory and inhibitory Poisson-train inputs at two locations on the dendrite. We first give either an excitatory or an inhibitory Poisson input alone with a duration of 1000 ms to measure the reference effective conductances from Eq. \ref{eqn:synaptic_G} in the absence of the current clamp. In the presence of an injected clamp current at the soma with various magnitudes, we then give simultaneously the excitatory and inhibitory Poisson inputs which are identical to those in the reference case. By recording the synaptic current and the membrane potential, we can obtain a linear I-V relation at each moment of time. By changing again the inhibitory reversal potential from $-10$ mV to $-20$ mV and following the same procedure above, we can obtain another linear I-V relation for the corresponding time. Using either the intercept method or the slope-and-intercept method, we can determine the corresponding excitatory and inhibitory conductances from the I-V relations. We now compare the values of the conductances measured by the two methods in the presence of the clamp current with those of the reference conductances measured in the absence of the clamp current. From Figure 3B, it is evident that the performance of the intercept method is significantly superior to the slope-and-intercept method. The effective conductances measured using the intercept method have a relatively small error --- with a time averaged error of 20\% for excitatory conductance and 10\% for inhibitory conductance (the origin of the error will be discussed in the \emph{Discussion} section). In contrast, those determined using the slope-and-intercept method yield a time averaged error as large as 25\% for excitatory conductance and 36\% for inhibitory conductance. To study the spatial effect, we next scan the input locations on the dendrite for a pair of excitatory and inhibitory input of constant synaptic conductances. For both excitatory and inhibitory inputs, the input site is scanned from 50 $\mu$m to 600 $\mu$m away from the soma. For each pair of input locations, we first measure the reference effective conductances in the absence of the injected clamp current. In the presence of the clamp current, we then determine the conductances using the intercept method in contrast to the slope-and-intercept method. As shown in Figures 3C-3F, the relative error of both methods increases as the stimulus location moves away from the soma to the distal dendrite. However, the intercept method produces relatively reliable results even for distal inputs with an error less than 10\%, whereas the error of the slope-and-intercept method can increase as drastically as to 60\%, giving rise to a rather unreliable value of conductance. We further note that the intercept method is also valid to determine conductance when voltage clamp is applied to the soma of the neuron. Figure S1 illustrates an example of the case of voltage clamp.
\begin{figure}
  \begin{center}
    \includegraphics[width=1\textwidth]{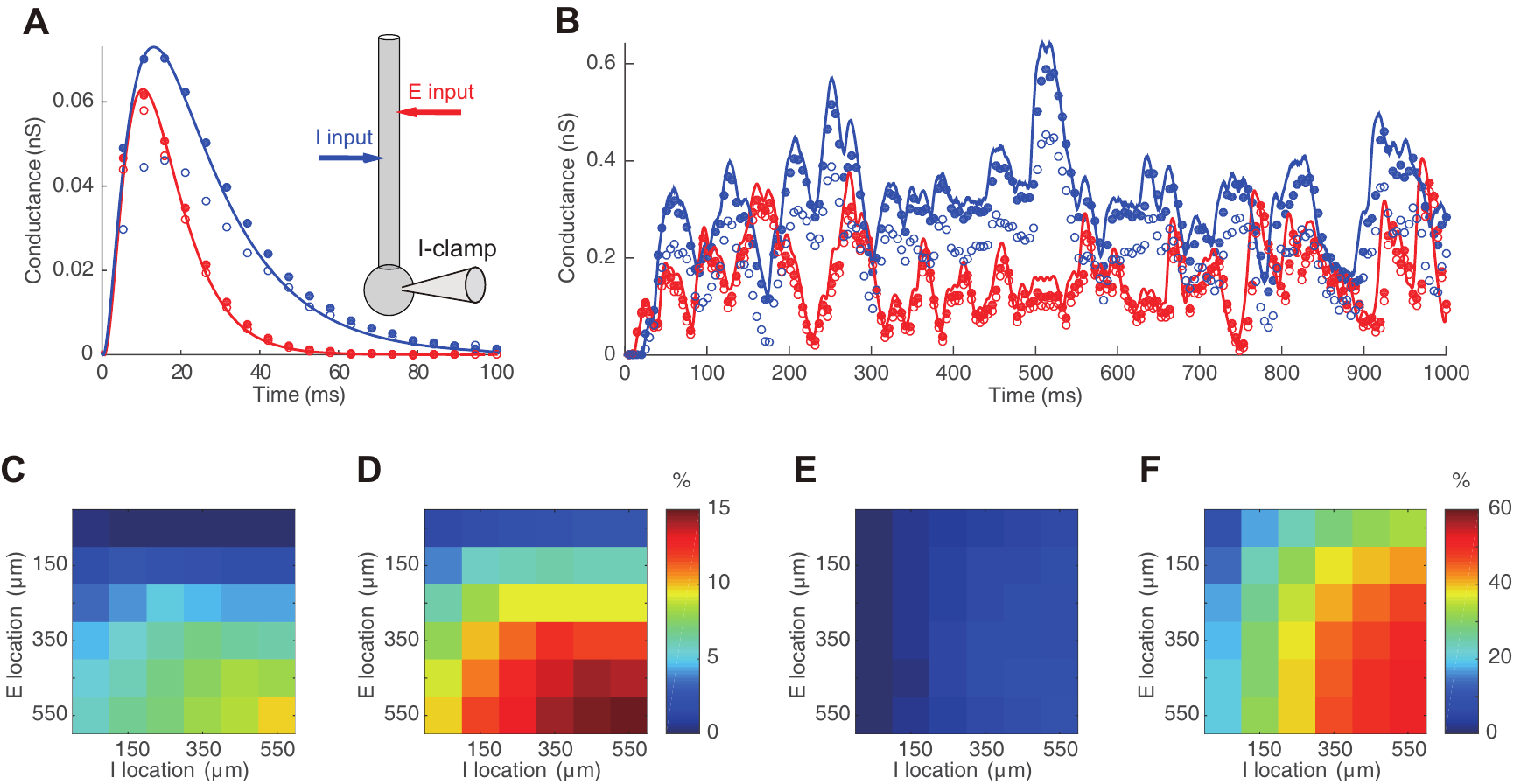}
  \end{center}
  \caption{Determination of effective conductance in the ball-and-stick neuron with current clamp at the soma. (A-B), the effective excitatory conductance $G^{\textrm{eff}}_{E}$ (solid red dots) and inhibitory conductance $G^{\textrm{eff}}_{I}$ (solid blue dots) determined by the intercept method nearly overlaps with the true value of the effective excitatory (red line) and inhibitory (blue line) conductances, whereas the excitatory conductance $G_{E}$ (open red circle) and the inhibitory conductance $G_{I}$ (open blue circle) determined by the slope-and-intercept method deviates from the true values. The deviation is particularly significant for the inhibitory case. (A) is for the case of paired transient excitatory and inhibitory pulse inputs. Inset in \emph{A} is the schematic diagram of the recording configuration. (B) is for the case of two excitatory and inhibitory Poisson inputs both with rate of 150 Hz. In \emph{A} and \emph{B}, excitatory and inhibitory inputs are given respectively at 420 $\mu$m and 300 $\mu$m away from the soma. (C-F) Spatial dependence of the relative error for the excitatory and inhibitory conductance measurement. Here, the location for a pair of excitatory and inhibitory input of constant conductances are scanned across the dendrites. The location distance is measured from the soma. (C-D) are the error of excitatory conductance measured by the intercept method and the slope-and-intercept method, respectively, with the same color bar to indicate the percentage of error. (E-F) are the error of inhibitory conductance measured by the intercept method and the slope-and-intercept method, respectively, with the same color bar to indicate the percentage of error. The excitatory and inhibitory reversal potentials relative to the resting potential are $\varepsilon_{E}=70$ mV, $\varepsilon_{I}=-10$ mV, respectively. The inhibitory reversal potential is changed to $\varepsilon_{I}=-20$ mV once while maintaining the excitatory reversal potential unchanged to obtain two I-V relations for the application of the intercept method.}
\end{figure}

Because a biological neuron possesses, in general, a complicated dendritic morphology with rich active ionic channels, it is necessary to use a realistic active neuron to further validate our method, as with the passive neuron model above. To address this, we deploy a biologically realistic pyramidal neuron model with tree-like dendritic morphology and broadly distributed active ionic channels. The data collection procedure is the same as that used in the case of the ball-and-stick neuron. Once more, we first numerically determine the value for a pair of transient conductance pulses as a reference true value when excitatory and inhibitory pulse inputs are given alone at locations on the dendritic trunk without an injected current at the soma. Now under the somatic current clamp mode, as shown in Figure 4A, the conductances measured using the intercept method have a relatively small error compared with the reference ones with a maximum value of 11\% for excitatory conductance and 6\% for inhibitory conductance. In contrast, those determined using the slope-and-intercept method yield an error as large as 33\% for excitatory conductance and 72\% for inhibitory conductance. To model the situation \emph{in vivo}, we distribute 15 excitatory inputs and 5 inhibitory inputs across the entire dendritic tree of the pyramidal neuron. At each synapse location, the arrival time of each input is randomly set between 0 ms and 1000 ms with input rate of 100 Hz. We use the identical input to measure time evolution of the reference effective conductances in the absence of the clamp current and to measure that of the conductances in the presence of the clamp current. We perform the comparison of the values of the conductances measured by the two methods with those of the reference conductances. As shown in Figure 4B, the excitatory and inhibitory conductances determined by our intercept method are, in general, close to the true value of the effective conductance except for the moments when the neuron generates action potentials. In contrast, the conductance determined by the traditional slope-and-intercept method deviates greatly from the true one, in particular, again for the inhibitory case. Under the subthreshold regime, the conductance measured by the intercept method has a relatively small error with time averaged error of 26\% for excitatory conductance and 13\% for inhibitory conductance. For comparison, those determined using the slope-and-intercept method yield a time averaged error as large as 51\% for excitatory conductance and 102\% for inhibitory conductance. In our numerical simulation, the true value of inhibitory conductance is significantly greater than that of the excitatory conductance. However, the inhibitory conductance estimated by the slope-and-intercept method is even smaller than the estimated excitatory conductance. Finally, as an illustration of the drastic failure of the traditional method, we stress that the inhibitory conductance can take an unphysical value to become negative, as can be clearly observed in Figure 4B.

In addition, as with the case of the ball-and-stick neuron, the measurement error induced by both methods is dependent of location. By scanning the location across the dendrite, for a pair of constant excitatory and inhibitory conductance inputs, we observe that the intercept method produces reasonably good results with an error of less than 10\% even for distal inputs, whereas the error of the slope-and-intercept method grows rapidly to 100\% as the input location moves away from the soma to the distal dendrites. From our simulation results, the conductance measured by the slope-and-intercept method is always smaller than the true effective conductance. Therefore, an error of 100\% corresponds to the case that the measured conductance vanishes while the true one is nonzero. The inhibitory conductance value measured by the slope-and-intercept method could again become negative at distal dendrites --- Figure 4F illustrates this severe problem with the traditional method. The intercept method is also verified to obtain correct conductance values when voltage clamp is applied to the soma of the realistic neuron, whereas the traditional slope-and-intercept method can produce negative conductance values, notably for inhibitory conductance (see Figure S2).

When a neuron generates an action potential, biologically, its membrane changes from the passive dynamics to an active one. The ionic current induced by the activation of voltage-gated ion channels can dominate effective synaptic currents induced by local synaptic inputs from the dendrite. As a consequence, the determination of the effective synaptic current based on Eq. \ref{eqn:synaptic_I} and thereby the corresponding effective synaptic conductance can have significant errors. This underlies the substantial difference, as observed in Figures 4B and S2B, between the conductance extracted by the intercept method and the reference conductance during an action potential generation. As pointed out by previous theoretical studies \cite{guillamon2006estimation}, it is rather difficult to determine conductance from the I-V relation when there is an action potential. To overcome this difficulty, in principle, one can pharmacologically block the active sodium channel to suppress the action potential generation. Thereby, the value of conductance measured by the intercept method approaches that of the reference conductance, which reflects the true strength of the effective conductance induced by synaptic inputs on the dendrites.

\begin{figure}
  \begin{center}
    \includegraphics[width=1\textwidth]{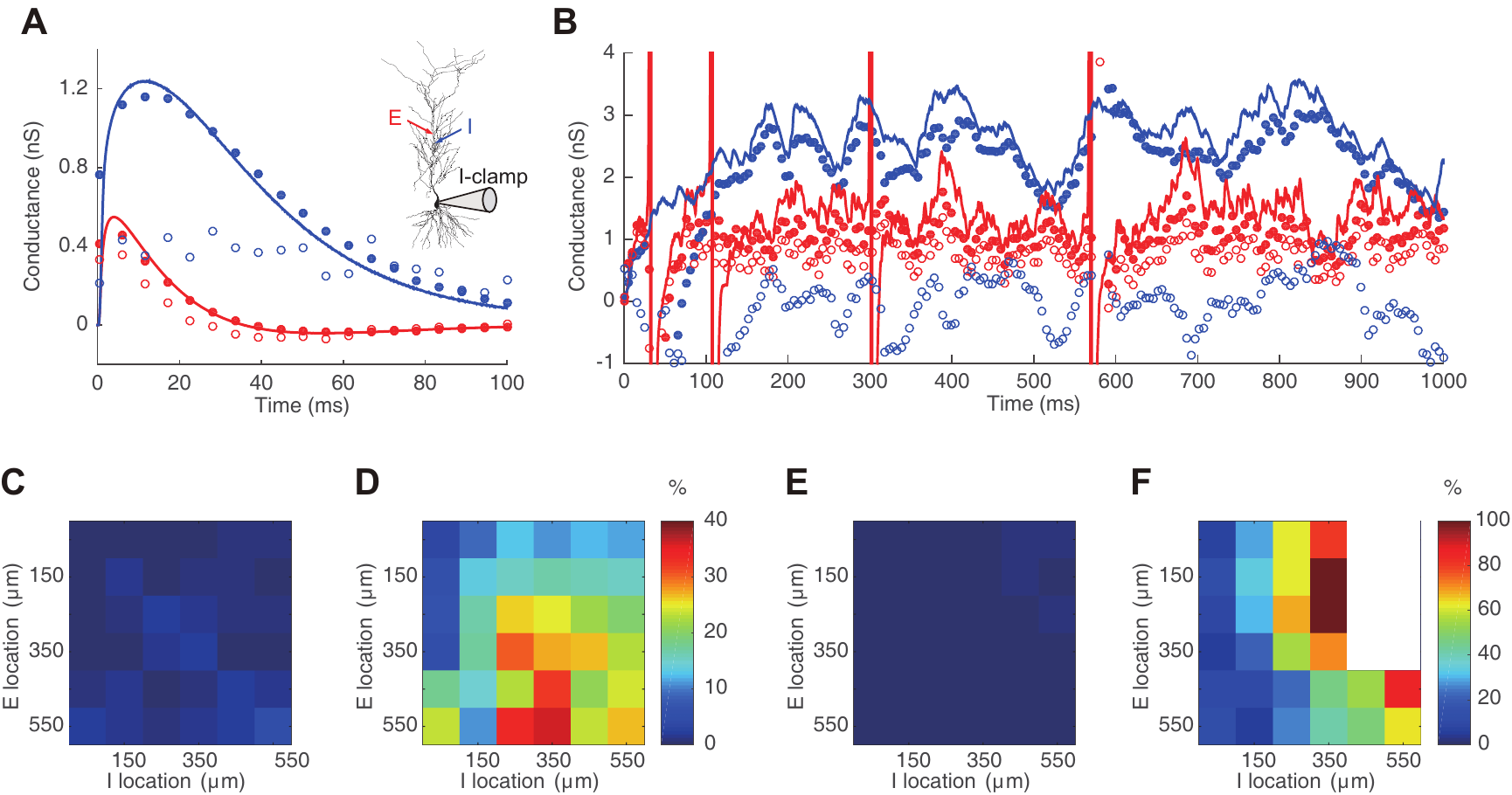}
  \end{center}
   \caption{Determination of effective conductance in the realistic pyramidal neuron model with current clamp at the soma. (A-B), the effective excitatory conductance $G^{\textrm{eff}}_{E}$ (solid red dots) and inhibitory conductance $G^{\textrm{eff}}_{I}$ (solid blue dots) determined by the intercept method are in good agreement with the true values of the effective excitatory (red line) and inhibitory (blue line) conductances, whereas the excitatory conductance $G_{E}$ (open red circle) and the inhibitory conductance $G_{I}$ (open blue circle) measured by the slope-and-intercept method deviate from the true values. The deviation is particularly significant for the inhibitory case. The failure of the traditional method is highlighted by the fact that the conductance measured by the traditional method can become negative. (A) is for the case of paired transient excitatory and inhibitory pulse inputs located on the dendrites 350 $\mu$m and 300 $\mu$m away from the soma. Inset in \emph{A} is the schematic diagram of the recording configuration. (B) is for the case of multiple inputs of 15 excitatory and 5 inhibitory locations distributed across the entire dendritic tree with rate of 100 Hz at each site. (C-F) Spatial dependence of the relative error for the excitatory and inhibitory conductance measurement. Here, the locations for a pair of excitatory and inhibitory inputs with constant conductances are scanned across the dendrites. The location distance is measured from the soma. (C-D) are the error of excitatory conductance measured by the intercept method and the slope-and-intercept method, respectively, with the same color bar to indicate the percentage of error. (E-F) are the error of inhibitory conductance measured by the intercept method and the slope-and-intercept method, respectively, with the same color bar to indicate the percentage of error. In \emph{F}, the large white area, which corresponds to negative conductance values, demonstrates the drastic failure of the traditional method. The excitatory and inhibitory reversal potentials relative to the resting potential are $\varepsilon_{E}=70$ mV, $\varepsilon_{I}=-10$ mV, respectively. The inhibitory reversal potential is again changed to $\varepsilon_{I}=-20$ mV once while keeping the excitatory reversal potential unchanged to obtain two I-V relations for the application of the intercept method.}
\end{figure}

\section*{Discussion}
In order to extract excitatory and inhibitory conductances, in many of previous works, a neuron with its complex dendritic arbor is assumed to be a single electrically compact compartment. Thereby, the somatic voltage clamp is deemed to uniformly control the membrane potential throughout the whole neuron. However, recent theoretical and experimental studies \cite{williams2008direct,poleg2011imperfect} have shown there is the space clamp effect, which limits the control of the voltage clamp on the membrane potential across the dendritic arbor. The membrane potential at distal synapses can deviate greatly from the holding potential, and the value of excitatory and inhibitory conductances measured in voltage clamp mode can be significantly distorted. In our work, we have made no attempt to eliminate or attenuate the space clamp effect. Instead, we have explored the possibility of viewing the soma rather than the whole neuron as an electrically compact point (Eq. \ref{eqn:newpoint}), so as to deploy the perfectly-clamped voltage at the soma to measure effective excitatory and inhibitory conductances at the soma, which contain the effect of the active ion channels along the dendrites and the filtering property of the dendrite. Conceptually, the effective conductance is a functionally important quantifier because they are strongly correlated to the local postsynaptic conductances at the dendrites and characterizes more directly than the local conductance the functional impact of synaptic inputs on the subthreshold dynamics and the spike trigger mechanism.

Using the traditional slope-and-intercept method, we have further shown that the measured conductance has no clear biological interpretation. It is close to neither the local synaptic conductance on the dendrite nor the effective synaptic conductance. In particular, we emphasize that the value of the conductance determined by the traditional method can be unphysically negative. Based on our theoretical analysis of the ball-and-stick neuron model, we have revealed that the failure of the slope-and-intercept method is caused by the interaction between the synaptic current and the clamp current. This interaction has not been addressed in previous studies. From our analysis, we have proposed a novel intercept method to measure the effective conductance accurately. We have verified the intercept method in both the numerical simulation of the ball-and-stick neuron model and the realistic pyramidal neuron model. Our numerical results show that, in general, the intercept method greatly improves upon the traditional slope-and-intercept method in current or voltage clamp by providing more reliable and accurate values of the effective excitatory and inhibitory conductances than the traditional method.

We note that the measured value of inhibitory conductance by the traditional slope-and-intercept method is much more distorted than that of excitatory conductance. This arises because the ratio of the excitatory to inhibitory conductance measurement errors is dependent of the excitatory reversal potential (70mV relative to the resting potential) and the inhibitory reversal potential (-10mV relative to the resting potential). We can analytically demonstrate that, the prefactors in the slope expression (Eq. \ref{eqn:slope}) in front of $G^{\textrm{eff}}_{E}$ and $G^{\textrm{eff}}_{I}$ can cause the measurement errors in the slope-and-intercept method amplified by a factor of $\varepsilon_{E}$ for $G_{I}$ and amplified by a factor of $\varepsilon_{I}$ for $G_{E}$. Therefore, the strength of inhibitory inputs can be significantly underestimated compared with excitatory inputs. One cannot simply compare the amplitudes of $G_{E}$ and $G_{I}$ measured from the traditional slope-and-intercept method to characterize network states such as the balanced state based on the relative amplitudes of $G_{E}$ and $G_{I}$.

In our work, the soma of the neuron is characterized by the leaky integrator (Eq. \ref{eqn:leaky}). If the synaptic input is so strong as to initiate action potentials, in order to measure the effective conductances, one can either pharmacologically block the action potential related active channels or switch the point neuron characterization from the integrate-and-fire type to others, for instances, the exponential integrate-and-fire neuron.

In our intercept method, conductance measurement error cannot be fully eliminated despite the above demonstration that our method produces rather good estimate of conductance in a biologically realistic neuron model, because we note that our analysis is accurate only to the first order approximation, higher order corrections may also contribute to the conductance value; Our analysis is based on the point-neuron model at the soma (Eq. \ref{eqn:newpoint}), however, the dendritic integration of synaptic inputs can lead to a more complicated form of the point-neuron model \cite{zhou2013phenomenological}. These are important issues for future studies. A precise understanding of the synaptic physiology of neurons using accurate effective conductances is important for investigating synaptic mechanisms of sensory processing, the origination of neuronal oscillations, and the balanced nature of excitation and inhibition in the brain.

\section*{Acknowledgments}
This work was supported by NYU Abu Dhabi Institute G1301 (S.L., D.Z., and D.C.), NSFC-11671259, NSFC-11722107, NSFC-91630208, Shanghai Rising-Star Program-15QA1402600 (D.Z.), NSFC 31571071, NSF DMS-1009575 (D.C.), Shanghai 14JC1403800, Shanghai 15JC1400104, SJTU-UM Collaborative Research Program (D.Z. and D.C.), the State Key Research Program of China 2011CBA00404 (X-h. Z.)\\

\noindent Correspondence and requests for materials should be addressed to D. Z. (zdz@sjtu.edu.cn), or D.C. (cai@cims.nyu.edu).
\begin{small}
\bibliographystyle{plain}
\bibliography{arXiv_template}
\end{small}

\section*{Supplementary Information}
\begin{figure}
  \begin{center}
    \includegraphics[width=1\textwidth]{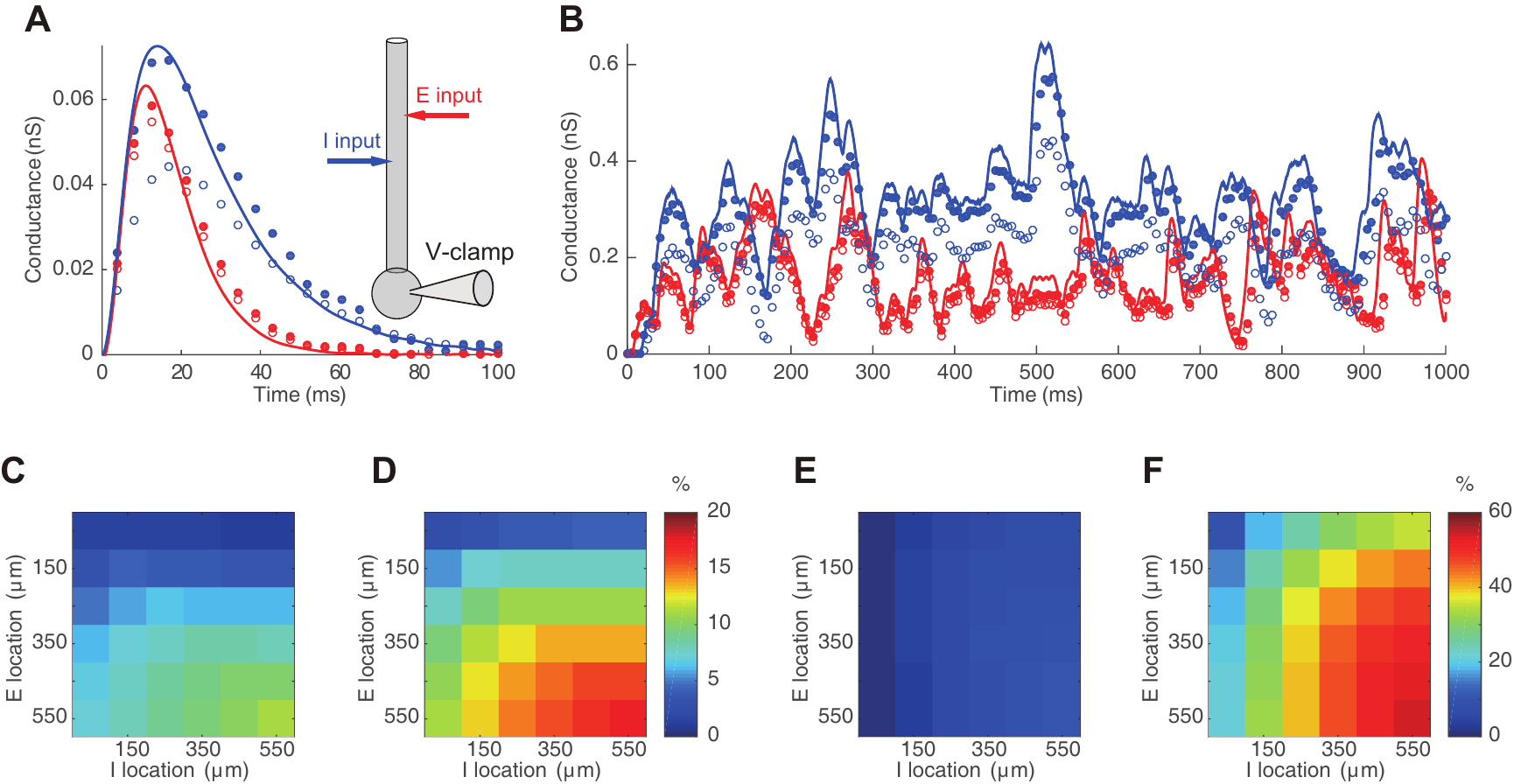}
  \end{center}
  \caption*{{\bf Figure S1.} Determination of effective conductance in the ball-and-stick neuron with voltage clamp at the soma. (A-B), the effective excitatory conductance $G^{\textrm{eff}}_{E}$ (solid red dots) and inhibitory conductance $G^{\textrm{eff}}_{I}$ (solid blue dots) determined by the intercept method is rather close to the true value of the effective excitatory (red line) and inhibitory (blue line) conductances, whereas the excitatory conductance $G_{E}$ (open red circle) and the inhibitory conductance $G_{I}$ (open blue circle) determined by the slope-and-intercept method deviates from the true values. As with the current clamp, the deviation is particularly significant for the inhibitory case. (A) is for the case of paired transient excitatory and inhibitory synaptic pulse inputs. Inset in \emph{A} is the schematic diagram of the recording configuration. (B) is for the case of two excitatory and inhibitory Poisson inputs both with rate of 150 Hz. In \emph{A} and \emph{B}, excitatory and inhibitory inputs are given respectively at 420 $\mu$m and 300 $\mu$m away from the soma. (C-F) Spatial dependence of the relative error for the excitatory and inhibitory conductance measurement. Here, the location for a pair of excitatory and inhibitory input of constant conductances is scanned across the dendrites. The location distance is measured from the soma. (C-D) are the error of excitatory conductance measured by the intercept method and the slope-and-intercept method, respectively, with the same color bar to indicate the percentage of error. (E-F) are the error of inhibitory conductance measured by the intercept method and the slope-and-intercept method, respectively, with the same color bar to indicate the percentage of error. The excitatory and inhibitory reversal potentials relative to the resting potential are $\varepsilon_{E}=70$ mV, $\varepsilon_{I}=-10$ mV, respectively. The inhibitory reversal potential is changed to $\varepsilon_{I}=-20$ mV once while maintaining the excitatory reversal potential unchanged to obtain two I-V relations for the application of the intercept method.}
\end{figure}

\begin{figure}
  \begin{center}
    \includegraphics[width=1\textwidth]{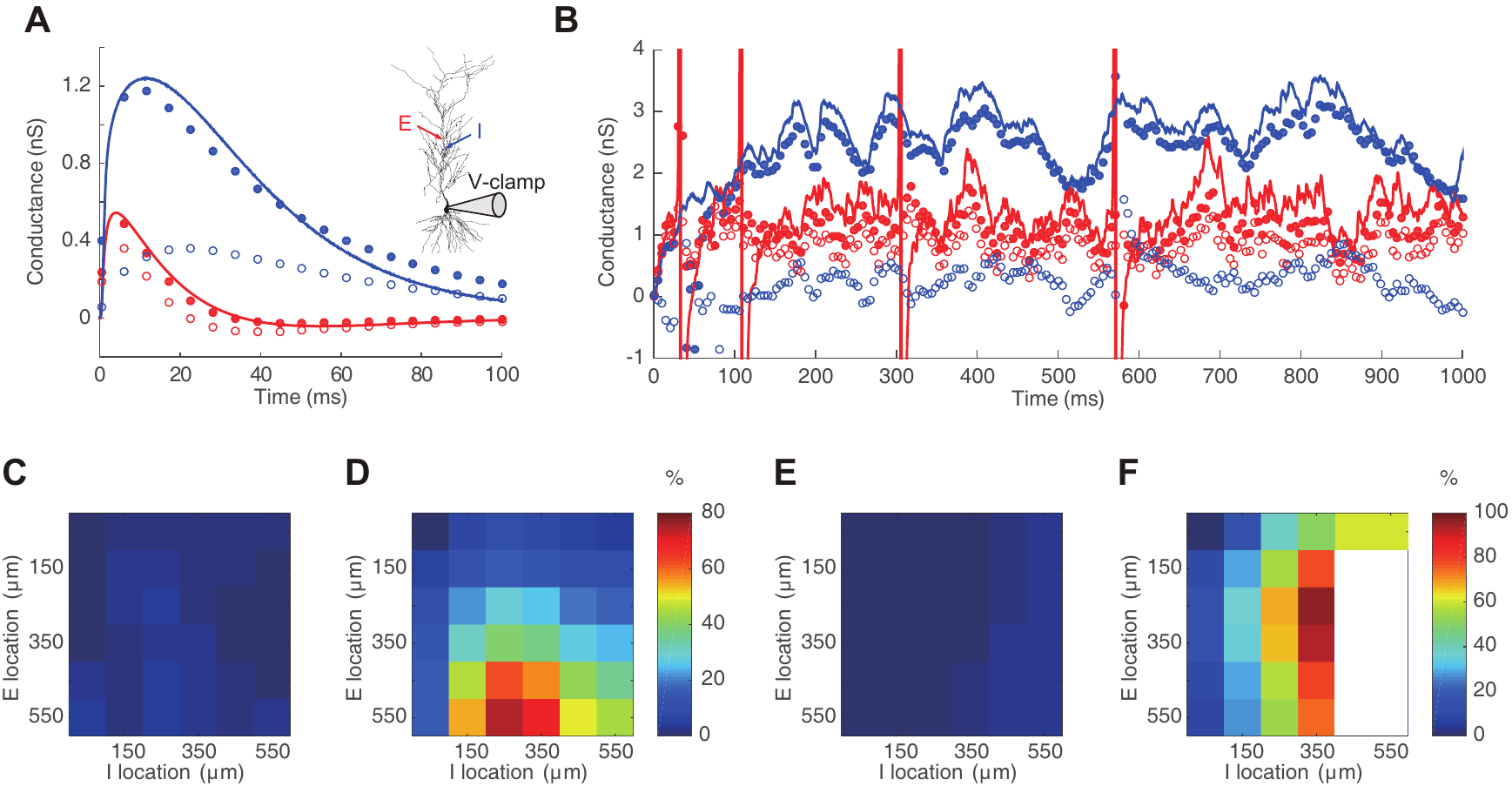}
  \end{center}
 \caption*{{\bf Figure S2.} Determination of effective conductance in the realistic pyramidal neuron model with voltage clamp at the soma. (A-B), the effective excitatory conductance $G^{\textrm{eff}}_{E}$ (solid red dots) and inhibitory conductance $G^{\textrm{eff}}_{I}$ (solid blue dots) determined by the intercept method are relatively close to the true values of the effective excitatory (red line) and inhibitory (blue line) conductances, whereas the excitatory conductance $G_{E}$ (open red circle) and the inhibitory conductance $G_{I}$ (open blue circle) measured by the slope-and-intercept method deviate greatly from the true values. The deviation is particularly significant again for the inhibitory case. The traditional method can even produce a negative value for conductance, again demonstrating the deficiency of the traditional method. (A) is for the case of paired transient excitatory and inhibitory pulse inputs located on the dendrites 350 $\mu$m and 300 $\mu$m away from the soma. Inset in \emph{A} is the schematic diagram of the recording configuration. (B) is for the case of multiple inputs of 15 excitatory and 5 inhibitory locations distributed across the entire dendritic tree with rate of 100 Hz at each site. (C-F) Spatial dependence of the relative error for the excitatory and inhibitory conductance measurement. Here, the locations for a pair of excitatory and inhibitory input of constant conductances are scanned across the dendrites. The location distance is measured from the soma. (C-D) are the error of excitatory conductance measured by the intercept method and the slope-and-intercept method, respectively, with a common color bar to indicate the percentage of error. (E-F) are the error of inhibitory conductances measured by the intercept method and the slope-and-intercept method, respectively, with a shared color bar to indicate the percentage of error. In \emph{F}, the large white area, which corresponds to negative conductance values, again illustrates the failure of the traditional method. The excitatory and inhibitory reversal potentials relative to the resting potential are $\varepsilon_{E}=70$ mV, $\varepsilon_{I}=-10$ mV, respectively. The inhibitory reversal potential is changed to $\varepsilon_{I}=-20$ mV once while keeping the excitatory reversal potential the same to obtain two I-V relations for the application of the intercept method.}
\end{figure}

\end{document}